# Semiclassical Monte Carlo Model for In-Plane Transport of Spin-Polarized Electrons in III-V Heterostructures

Semion Saikin[a,b,c], Min Shen[a], Ming-C. Cheng[a] and Vladimir Privman[a,b]

[a] Center for Quantum Device Technology,

[a] Department of Electrical and Computer Engineering and

[b] Department of Physics, Clarkson University,

Potsdam, New York 13699-5720, USA

[c] Department of Physics,

Kazan State University, Kazan, Russian Federation

**ABSTRACT**

We study the in-plane transport of spin-polarized electrons in III-V semiconductor quantum wells. The spin dynamics is controlled by the spin-orbit interaction, which arises via the Dresselhaus (bulk asymmetry) and Rashba (well asymmetry) mechanisms. This interaction, owing to its momentum dependence, causes rotation of the spin polarization vector, and also produces effective spin dephasing. The density matrix approach is used to describe the evolution of the electron spin polarization, while the spatial motion of the electrons is treated semiclassically. Monte Carlo simulations have been carried out for temperatures in the range 77-300 K.

# I. INTRODUCTION

Promising applications of spintronics for novel device structures[1-5] have stimulated much interest in spin polarized transport. Many devices utilizing spin-dependent phenomena have been proposed recently.[6-14] At the present time, there are numerous difficulties with control of spin polarized current. Recent experimental advances[2] have allowed generation of spin polarization of conduction electrons in bulk semiconductors and in two-dimensional semiconductor structures. At room temperatures, spin polarization can be maintained for up to 1-2 nanoseconds.

Experimental investigations of spin polarized transport in semiconductors can be divided into three main areas: injection and detection of spin polarized current, spin relaxation of conduction electrons, and coherent spin dynamics.

Among few methods to create electron spin polarization in semiconductors,[15-18] the electrical spin injection from magnetic contacts[17,18] is the most promising. However, the main difficulty of this approach has been in the correct band matching at the interface of magnetic material – semiconductor.[19,20] Also, the all-electrical experiments on the detection of spin injection are complicated by additional spin independent effects, which are difficult to separate from spin-dependent phenomena.[21] Thus, the recently reported values of the experimentally achieved spin polarization at room temperature have varied from 1-2%[22] up to 30-35%.[23,24] At low temperatures, $T \sim 4.2$ K, the polarization of the electrons injected from magnetic semiconductor contacts is appreciably higher and reaches the values of 50-80%.[25,26]

The optical electron spin polarization and detection methods[16] are, perhaps, less applicable in the device design, but they have been very useful in investigations of electron spin dynamics, due to high efficiency of spin polarization at room temperature (more than 50%) and high sensitivity of measurement. Spin relaxation in semiconductor heterostructures has been studied extensively by the methods of ultrafast spin-sensitive spectroscopy.[27-31] At room temperature, the observed spin relaxation time varies widely, from less than 1 ps for structures with large spin-orbit interaction,[27] up to 1 ns [28] for GaAs/AlGaAs quantum wells (QW) with suppressed Dyakonov-Perel relaxation mechanism.[32]

The spin-lattice relaxation of conduction electrons at high temperatures is dominated by several mechanisms arising from spin-orbit coupling. Their relative strength is determined by many different factors, some of which are established during the growth of the heterostructure and can not be well controlled. While for n-doped GaAs/AlGaAs QWs with (001) growth orientation the main relaxation mechanism at room temperature is Dyakonov-Perel,[30] spin relaxation in narrow band gap (InGaAs/InP, InGaAs/InAlAs) heterostructures has no single explanation due to more complicated spin-orbit interactions.[29,31]

Generally, the conduction electron spin dynamics is controlled by external magnetic field, local magnetic fields produced by magnetic impurities and nuclei, and spin orbit interactions. In comparison with the electron spin transport model for ferromagnetic structures, which can be described within the two-current model,[33] for nonmagnetic bulk semiconductors and semiconductor heterostructures the spin-orbit term is significant. This effect has been investigated by analyzing the Shubnikov-de Haas oscillations of the magnetoresistance in moderate magnetic fields[34,35] and studying weak antilocalization in nearly-zero fields.[36-38] The tuning of the spin-orbit coupling constant by the gate voltage, which has been proposed for current modulation in the spin-FET,[6] was demonstrated recently for InGaAs/InAlAs asymmetric QWs.[35,39]

In the low-temperature and low-voltage regime, the value of the spin mean free path in bulk GaAs can reach few μm,[40,41] which is much larger than industrially achievable device sizes.[42] The above overview of selected promising experimental results for spintronics device development, suggests that it is timely to develop device-modeling approaches incorporating spin polarization effects.

For low temperatures and low applied voltage, the single-particle ballistic models have been utilized.[13,43,44] Many of the existing semiclassical models have been developed along the lines of the earlier approach used for ferromagnetic layered structures.[33] They are primarily of the drift-diffusion category, where the spin-up and spin-down electrons are described by charge- and spin-density conservation equations.[45-47] These models ignore quantum coherence effects of possible superposition of the spin-up and spin-down states, which can be described in terms of the polarization vector.[48] The range of applicability for such models is limited by many factors (electric field, device size, etc.), when non-linear effects become important.[20,41] For hot-electron spin-polarized transport, the coupled Boltzmann equations with only spin-up and spin-down states,[49] or additionally including superimposed up-down states,[50,51] have been considered.

Spin relaxation of conduction electrons and their spatial motion can not be separated exactly. However, in some drift-diffusion approximations, it can be shown that spin polarization of the electron gas decays exponentially in time in accordance with the spin-relaxation Bloch equations[52,53] with characteristic times $T_1$ and $T_2$. It is reasonable to assume, in analogy with the energy and momentum relaxation times[54] in energy-balance or hydrodynamic models for semiconductor devices, that these parameters will depend on the electron temperature, $T_e$, and, possibly, some other variables.[55] Monte Carlo simulation including the electron spin state[56-58] can be useful for spin-dynamics modeling in the non-linear regime and extraction of such parameters. In this work, we utilize the Monte Carlo approach to simulate spin polarized transport in asymmetric QWs for intermediate values of the electric field (~2-4 kV/cm), for temperature $T$ = 77-300 K.

## II. SEMICLASSICAL DENSITY MATRIX APPROACH TO SPIN POLARIZED ELECTRON TRANSPORT

Monte Carlo approach to Boltzmann equation for non-stationary electron transport has been widely used for modeling of submicrometer and deep-submicrometer devices.[59,60] Here, we incorporate the description of the electron spin dynamics in a standard semiclassical Monte Carlo formalism.[59,60] The effective single-electron Hamiltonian with the spin-orbit interaction term is

$$H = H_0 + H_{SO}(\boldsymbol{\sigma}, \mathbf{k}) \quad . \tag{1}$$

In the absence of external and local magnetic fields, the electron magnetic interaction is only owing to the spin-orbit term $H_{SO}(\boldsymbol{\sigma}, \mathbf{k})$ in Eq. (1). $H_0$ is the self-consistent single electron Hamiltonian in the Hartree approximation, including also interactions with phonons and static imperfections. Inside the QW, this can be written as

$$H_0 = -\frac{\hbar^2}{2m^*}k^2 + e\gamma(\mathbf{r})(\mathbf{E}_{ext} \cdot \mathbf{r}) + H_{e\text{-ph}} + H_{ph} + V_{imp} \quad . \tag{2}$$

In the semiclassical treatment,[59] the operator **k** is considered as the momentum vector in the *xy* plane of the QW, while the motion in the *z* direction is quantized. We consider here a structure grown in (0, 0, 1) direction, and assume that the in-channel electric field is applied along the *x* crystallographic axis. These assumptions allow us to specify the form of spin-orbit interaction term, Eq. (1).[32] The coordinate axes are chosen parallel to the crystallographic directions. The screening factor $\gamma(\mathbf{r})$ accounts for the electron-electron interactions. It is determined by the appropriate Poisson equation.[61] The term $V_{imp}$ describes ionized nonmagnetic impurities, QW roughness and other static imperfections of its structure. The terms labeled "e-ph" and "ph" represent the electron-phonon interactions and the phonon mode Hamiltonian. The main contributions to the spin-orbit interaction in an asymmetric III-V semiconductor QW structure are due to the Dresselhaus mechanism,[62,32]

$$H_D = \beta \langle k_z^2 \rangle (k_y \sigma_y - k_x \sigma_x) \ , \tag{3}$$

and Rashba mechanism,[63]

$$H_R = \eta (k_y \sigma_x - k_x \sigma_y) \ . \tag{4}$$

Equation (3) as written, is only applicable for narrow QWs, such that $k_x, k_y \ll \sqrt{\langle k_z^2 \rangle}$. For submicrometer devices with smooth potential, in the considered temperature regime (*T* = 77-300 K), we assume that the spatial electron transport is semiclassical and can be described by Boltzmann equation.[61] The electrons follow classical localized trajectories between the scattering events. The scattering rates are given by Fermi Golden Rule, and the scattering events are instantaneous. We also assume that the Elliott-Yafet spin scattering mechanism[64] is inefficient, i.e., that there are no electron spin flips

- 6 -

accompanying momentum scattering. The back reaction of the electron spin evolution on the spatial motion is negligible owing to the small value of the electron momentum-state splitting due to spin-orbit interaction in comparison with its average momentum.

In Monte Carlo simulations, it is assumed that electrons propagate with constant velocity during the time $\delta t$, which is the smaller of the grid time step and the time interval either left to the next scattering or from scattering to the next sampling. We term such a motion "free flight." The propagation velocity of an electron in the "free flight" was taken as the average value of the velocity of an electron moving with constant acceleration during $\delta t$. Among many different scattering mechanisms,[65] our Monte Carlo simulation has included charged impurity and phonon scatterings. The phonon bath in Eq. (2) is assumed to remain in thermal equilibrium with the constant lattice temperature $T$ at all times. In the semiclassical Monte Carlo, the temperature is incorporated in the electron-phonon scattering rates.[59,60] Details of the Monte Carlo simulation model are described elsewhere.[66]

For the description of the electron spin, we use the standard spin density matrix,[67]

$$\rho_\sigma(t) = \begin{pmatrix} \rho_{\uparrow\uparrow}(t) & \rho_{\uparrow\downarrow}(t) \\ \rho_{\downarrow\uparrow}(t) & \rho_{\downarrow\downarrow}(t) \end{pmatrix}, \qquad (5)$$

which is associated with the spin polarization vector as $S_\zeta(t) = Tr(\sigma_\zeta \rho_\sigma(t))$, where $\sigma_\zeta$ ($\zeta = x, y, z$) are Pauli matrixes[67]. For each "free flight" time interval, $\delta t$, the spin density matrix evolves according to

$$\rho_\sigma(t + \delta t) = e^{-iH_{SO}\delta t/\hbar} \rho_\sigma(t) e^{iH_{SO}\delta t/\hbar}. \qquad (6)$$

This is equivalent to rotation of the spin polarization vector about the effective magnetic field determined by the direction of the electron momentum. The exponential operators in Eq. (6) can be written as (2×2) scattering matrices,

$$e^{-iH_{SO}\delta t/\hbar} = \begin{pmatrix} \cos(|\alpha|\delta t) & i\frac{\alpha}{|\alpha|}\sin(|\alpha|\delta t) \\ i\frac{\alpha^*}{|\alpha|}\sin(|\alpha|\delta t) & \cos(|\alpha|\delta t) \end{pmatrix}, \qquad (7)$$

and Hermitean conjugate of Eq. (7) for the operator $e^{iH_{SO}\delta t/\hbar}$. Here $\alpha$ is determined by the spin-orbit interaction terms, Eqs. (3,4),

$$\alpha = \hbar^{-1}\left[\left(\eta k_y - \beta\langle k_z^2\rangle k_x\right) + i\left(\eta k_x - \beta\langle k_z^2\rangle k_y\right)\right]. \qquad (8)$$

During the "free flight," the spin dynamics of a single electron spin is coherent; see Eq. (6). However, stochastic momentum fluctuations during the scattering events, produce distribution of spin states, thus causing effective dephasing at times $t > 0$. The spin polarization, $\langle S_\zeta\rangle$, of the current can be obtained by averaging $S_\zeta$ over all the electrons in a small volume $dv$, which is located at position $\mathbf{r}$, at time $t$. The absolute value of the average spin polarization vector is in the range $|\langle \mathbf{S}\rangle| \leq 1$. If $|\langle \mathbf{S}\rangle|$ is 1, the electric current is completely spin polarized. The components $\langle S_\zeta\rangle$ define the orientation of the spin polarization, and evolution of the spin polarization vector may be viewed as coherent motion (rotation) accompanied by depolarization (reduction of magnitude).

## III. MODEL AND SIMULATION RESULTS

We have utilized the model of spin polarized current, described in the preceding section, in simulation of electron transport in a single QW. Here, we utilize the asymmetric QW architecture in the one-subband approximation, which is a simplified model of $In_{0.52}Al_{0.48}As/In_{0.53}Ga_{0.47}As/In_{0.52}Al_{0.48}As$ heterostructures used in experiments probing the spin-orbit coupling effects.[35] Parameters of the confining potential, and spin-orbit coupling constants, used in our simulations, are given in Table I; see Fig. 1 for the potential shape. The Rashba and Dresselhaus coupling constants have been adopted from the literature.[35,68]

We denote by $n$ the equilibrium electron density in the channel. The ratio of the expectation values for the Dresselhaus and Rashba energy terms, Eqs. (3,4), is $E_R/E_D \approx 5.3$, which means that the Rashba term is dominant both for the coherent polarization-rotation dynamics and for depolarization. In our simulations, the device length was taken $l = 0.55\,\mu m$. The grid time step was $\delta t_{grid} = 1\,fsec$. The material and scattering-rate parameters were taken from the literature.[65] To achieve the steady-state transport regime, we ran the simulation program for 20000 time steps, and collected data only during the last 2000 time steps. The simulations were carried out for temperatures $T$ = 77-300 K and applied drain-source voltage $V_{DS}$ = 0.1-0.25 V, which creates the in-channel electric field of the order of 2-4.5 kV/cm. The following boundary conditions were assumed: thermalized electrons were injected at the left boundary, with 100% injected spin polarization, and drained at the right boundary, with any spin polarization.

**TABLE I.** Parameters of the confining potential, and the spin-orbit interaction coupling constants.

| $d$, nm | $\Delta E_c$, eV | $n$, cm$^{-2}$ | $E_1$, eV | $\langle k_z^2 \rangle^{1/2}$, nm$^{-1}$ | $\eta$, eV·Å | $\beta$, eV·Å$^3$ |
|---|---|---|---|---|---|---|
| 20 | 0.56 | $1\times10^{12}$ | 0.20 | 0.21 | 0.074 | 32.20 |

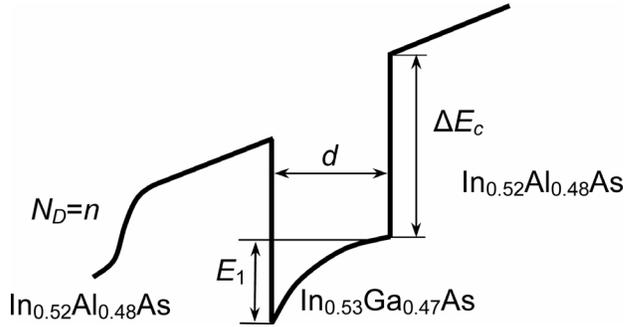

**FIG. 1.** Model of the confining potential in the asymmetric In$_{0.52}$Al$_{0.48}$As/In$_{0.53}$Ga$_{0.47}$As/In$_{0.52}$Al$_{0.48}$As quantum well.

In the simulated device structure, the electron transport is non-equilibrium; see Fig. 2. Evident velocity overshoot and other sharp features are observed due to sudden increase in the electric field near the injecting boundary at all applied voltages. The electron average energy in the two-dimensional quantum well includes the drift and thermal energies,

$$\langle E \rangle = \frac{1}{2} m^* \langle \mathbf{v} \rangle^2 + kT_e \quad . \tag{9}$$

Near the boundary at $x = 0$, where electrons are just injected, the thermal energy $kT_e$ is dominant.

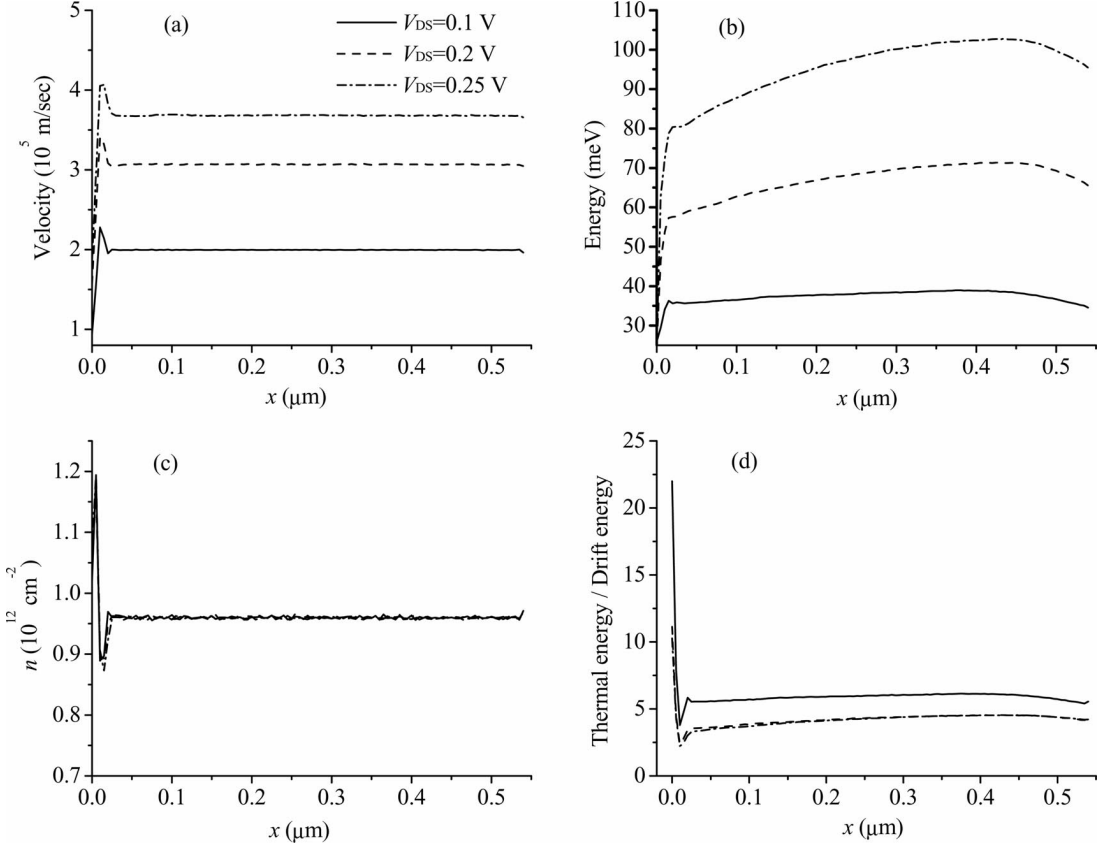

**FIG. 2.** The electron transport parameters: (a) drift velocity, (b) average energy, (c) electron concentration in the channel, and (d) electron thermal energy, as compared to their drift energy, as functions of $x$, at $T = 300$ K, $V_{DS} = 0.1$-$0.25$ V.

Due to finite scattering rate ($\sim 10^{-13}$ sec$^{-1}$), ballistic motion is observed in Fig. 2(a) for distances as small as 0.01μm, where average velocity increases considerably. This results in a sudden decrease in the ratio between the thermal and drift energies, as seen in Fig.

2(d), and leads to an abrupt increase in the average energy, shown in Fig. 2(b). After the ballistic region, electrons suffer strong scattering that randomizes momentum and gives rise to velocity overshoot. In order to maintain the current continuity, electron concentration markedly drops at the location of the velocity overshoot; see Fig. 2(c).

We calculate the evolution of the current spin polarization for three injected polarizations: along the positive $x$, $y$ and $z$ directions. The corresponding injected single-electron density matrixes are,

$$\rho_x(0) = \frac{1}{2}\begin{pmatrix} 1 & 1 \\ 1 & 1 \end{pmatrix}, \quad \rho_y(0) = \frac{1}{2}\begin{pmatrix} 1 & -i \\ i & 1 \end{pmatrix}, \quad \rho_z(0) = \begin{pmatrix} 1 & 0 \\ 0 & 0 \end{pmatrix}. \quad (10)$$

Due to the symmetry of the Rashba spin-orbit interaction term, Eq. (4), the spin polarization of the electrons, which propagate collectively in the $x$ direction, will rotate about the $y$ axis. This is shown in Fig. 3, where, from now on, we omit the angular brackets that indicate averaging. The Dresselhaus term, Eq. (3), causes rotation about the $x$ axis. Small admixture of the later mechanism leads to variation of the $y$ projection of the spin polarization, see Figs. 3(a), 3(c), and of the $x$ and $z$ projections, see Fig. 3(b), depending on the injected spin orientation.

The observed decay of the spin polarization occurs by dephasing owing to the electron momentum scattering events. Random momentum fluctuations, which are described by the electron thermal energy, produce an effective depolarization mechanism. The initial spin polarization drop in Fig. 3, can be attributed to the effect of high electron thermal energy in comparison with the drift energy, see Fig. 2(d). This can be clearly observed in Fig. 4(c), where the drop of the spin polarization near $x = 0$ is smaller for lower temperatures.

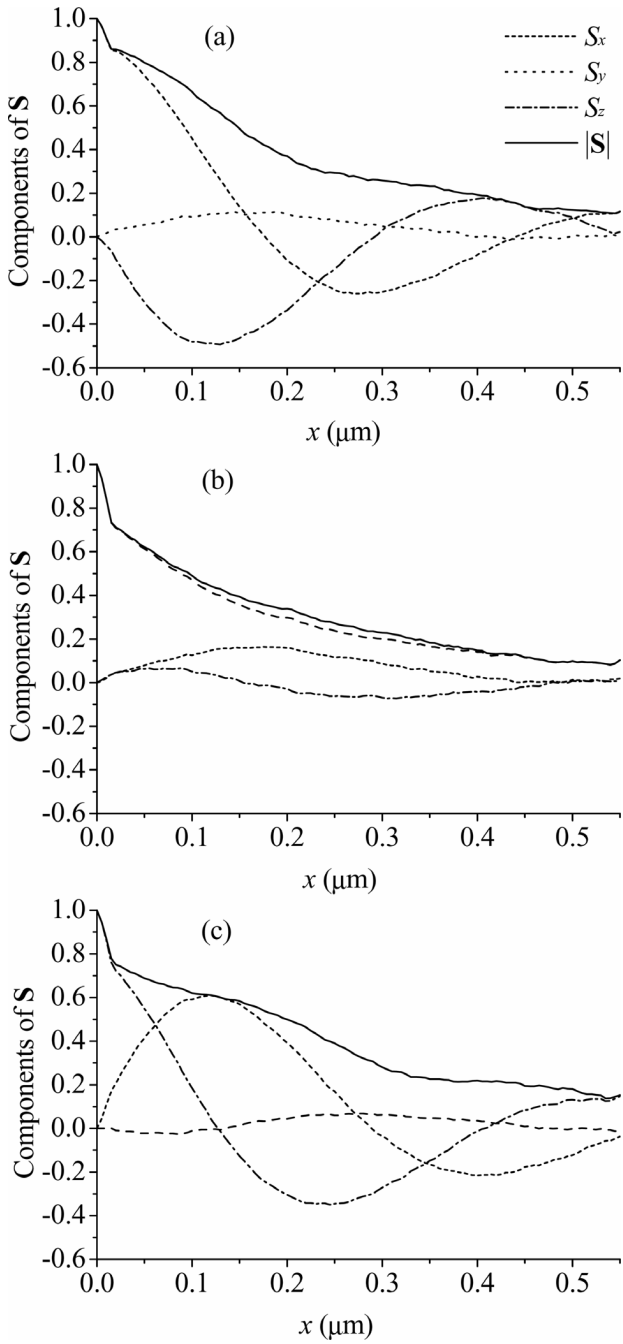

**FIG. 3.** Evolution of the electron spin polarization **S**, for $V_{DS} = 0.1$ V, at room temperature (300 K), for three different injected orientations of the spin polarization, along the positive (a) $x$, (b) $y$, and (c) $z$ axes.

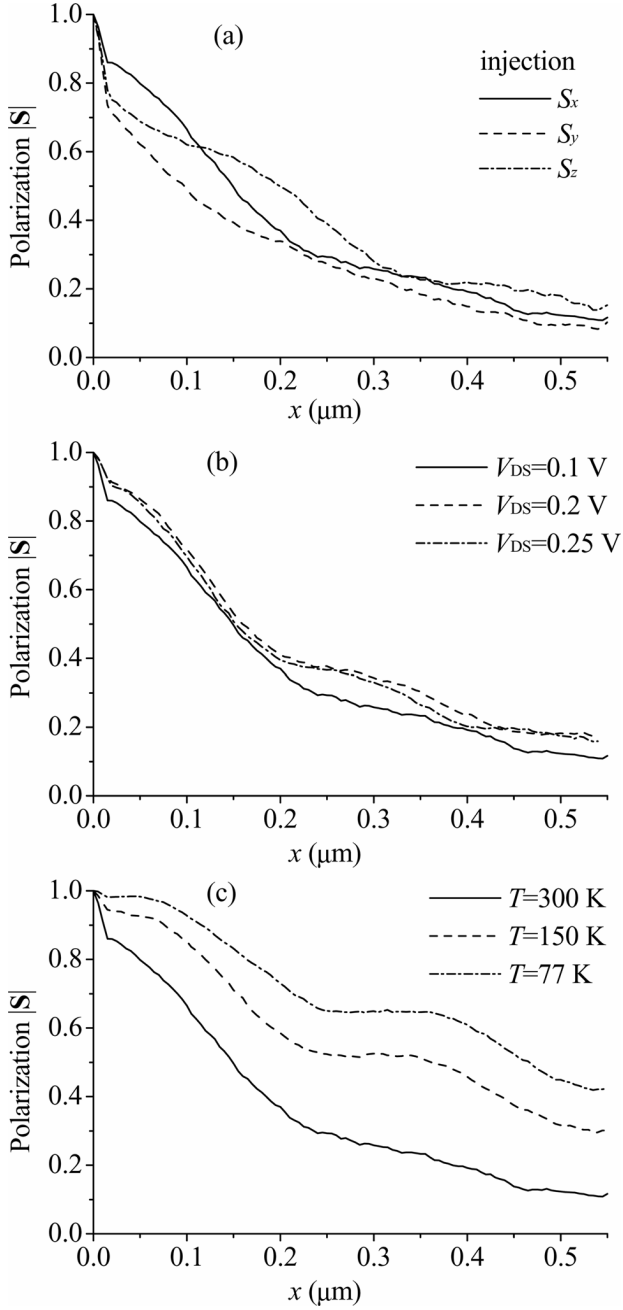

**FIG. 4.** Spin depolarization effect for (a) different orientations of the injected polarization, $T = 300$ K, $V_{DS} = 0.1$ V; (b) different values of applied voltage, $T = 300$ K, injected polarization $S_x = 1$; (c) different temperatures, injected polarization $S_x = 1$, $V_{DS} = 0.1$ V.

Moreover, Fig. 4(b) shows that this drop is evidently less pronounced at higher applied voltage that enhances the drift velocity and relatively weakens the effect of the random momentum fluctuations. This drop could also be an artifact of the "free flight" assumption for regimes with strong acceleration.

The depolarization rate in our model is asymmetric in the spin orientation. For example, the term proportional to $k_y \sigma_x$ in the Rashba spin-orbit interaction, Eq. (4), produces the depolarization of the $S_y$ and $S_z$ components, due to the fluctuating $k_y$,[6] but does not influence the $S_x$ component of the spin polarization. The depolarization rate owing to the Rashba interaction is suppressed for an $S_x$-polarized current in comparison with $S_y$ and $S_z$. This effect can be seen in Fig. 4(a).

As shown in Fig. 4(b), the spin polarization at room temperature is not sensitive to the applied voltage in the investigated regime. Higher applied voltage, which leads to considerably larger drift velocity, only slightly increases the spin-dephasing length. Change of the spin polarization at higher applied voltage is minimized by the increase in the mean energy that in turn increases the scattering probability. For lower temperatures, this quasi-balance apparently breaks down. Sufficient reduction of the temperature can suppress the electron-phonon scattering mechanism to yield longer spin mean free path. The temperature effect on the spin polarization in the range $T = 77$-$300$ K is shown in Fig. 4(c). The calculated values of the spin mean free path at $V_{DS} = 0.1$ V are $L_x \sim 0.2$ μm and $L_x \sim 0.55$ μm, for $T = 300$ K and $T = 77$ K, respectively. These values are significantly smaller than those obtained for bulk GaAs in the low-temperature ($T \sim 9$ K) regime,[40,41] $L_x > 4$ μm, which could be attributed to stronger scattering.

## IV. DISCUSSION

The simulation model developed in this paper, includes the linear terms of the spin-orbit coupling, which determine the spin energy basis. The terms cubic in the components of the momentum **k** in the Dresselhaus spin-orbit interaction,[62] and external magnetic field, can produce additional spin dephasing.[51,69] The Elliot-Yafet spin-scattering mechanism can be efficient in narrow-gap heterostructures. It can be included in our simulation model as an additional spin-evolution process at momentum scattering events.

The single subband approximation can be questioned for considered values of the applied voltage. For the confining potential used in the simulation, the estimated splitting between the ground and first-excited subbands is $\Delta E_{12} \sim 60 - 70$ meV. The intersubband scattering becomes effective at kinetic energies near $\Delta E_{12} - \hbar\omega_{LO} \sim 35$ meV, when optical phonon absorption becomes possible. Strong scattering sets in when the electron energy is above $\Delta E_{12} + \hbar\omega_{LO} \sim 100$ meV, and emission of optical phonons becomes possible. According to Fig. 2(b), the average electron energy exceeds the minimum value for the intersubband scattering for $V_{DS} > 0.1$ V. Up to the energy value of 100 meV, intersubband scattering rate is much less than intrasubband scattering rate,[61] and we can assume that the single subband approximation gives a qualitatively correct description up to the $V_{DS}$ = 0.25 V. For low temperatures, the average energy is reduced due to the condition of initial thermalization, Fig. 5(a), but, on the other hand, the electron-phonon scattering is suppressed, resulting in the energy increase, see Fig. 5(b). In comparison with the data shown in Fig. 2(b), the maximum value of the average electron energy for $V_{DS} = 0.2$ V is nearly the same. As a result, it is difficult to extend the validity of the single subband model for $V_{DS} > 0.2$ V by reducing the temperature. For different subbands, the spin-coupling constants are different, and it is likely that the spin dephasing will be even stronger if the intersubband scattering is incorporated.

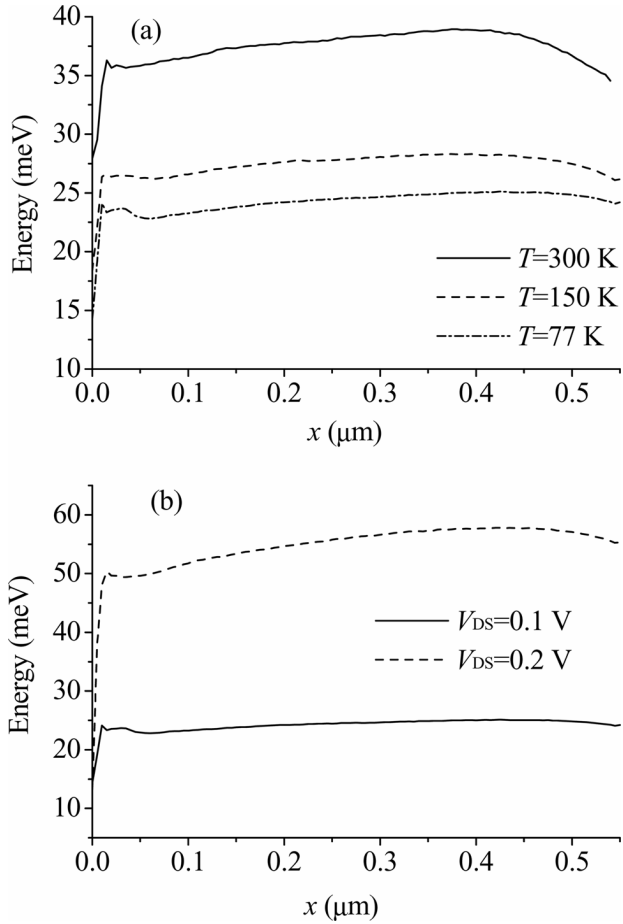

**FIG. 5.** The average electron energy for different values of the (a) temperature (at $V_{DS} = 0.1$ V); (b) applied voltage (at $T = 77$ K).

## V. SUMMARY

We have developed a semiclassical Monte Carlo model incorporating the linear terms of the Dresselhaus and Rashba spin-orbit coupling mechanisms for spin-polarized electron transport in III-V heterostructures. This approach can be used for simulation of non-equilibrium spin-dependent phenomena in spintronics devices. We reported results for dynamics of the spin polarization in a single quantum well at several temperatures and intermediate, ~ 2-4 kV/cm, electric fields. The estimated spin depolarization length is of the order of 0.2 μm. The present-day semiconductor device component dimensions are comparable or smaller. Thus, our results confirm that spintronic effects can be observed and controlled in properly designed modern semiconductor structures.

**ACKNOWLEDGMENTS**

We thank Professors A. Shik and I. D. Vagner for helpful discussions. This research was supported by the National Security Agency and Advanced Research and Development Activity under Army Research Office contract DAAD-19-02-1-0035, and by the National Science Foundation, grants DMR-0121146 and ECS-0102500.